\documentclass[final,5p,times,twocolumn]{elsarticle}

\usepackage{amssymb}
\usepackage{amsmath}
\usepackage{braket}
\usepackage{bbold}
\usepackage{hyperref}
\usepackage{multirow}
\newcommand{\bsxi}{\boldsymbol{\xi}}

\renewcommand{\vec}[1]{\boldsymbol{#1}}
\newcommand{\be}{\begin{equation}}
\newcommand{\ee}{\end{equation}}
\newcommand{\bea}{\begin{eqnarray}}
\newcommand{\eea}{\end{eqnarray}}
\newcommand{\Tr}{\,\hbox{\rm Tr}}

\def\MSbar{\overline{\rm MS\kern-0.5pt}\kern0.5pt}

\def\bsx{{\boldsymbol x}}
\def\bsxi{{\boldsymbol \xi}}

\begin{document}

\begin{frontmatter}

\title{Baryonic screening masses in QCD at high temperature}

\author[a,b]{Leonardo Giusti}
\author[c]{Tim Harris}
\author[a,b]{Davide Laudicina}
\author[b]{Michele Pepe}
\author[a,b]{Pietro Rescigno}

\affiliation[a]{organization={Department of Physics, University of Milano-Bicocca},
                addressline={Piazza della Scienza 3},
                postcode={20126},
                city={Milano},
                country={Italy}}
                
\affiliation[b]{organization={INFN, sezione di Milano - Bicocca},
                addressline={Piazza della Scienza 3},
                postcode={20126},
                city={Milano},
                country={Italy}}

\affiliation[c]{organization={Institut fur Theoretische Physik, ETH Zurich},
                addressline={Wolfgang-Pauli-Str. 27},
                city={Zurich},
                postcode={8093},                
                country={Switzerland}}

\begin{abstract}
We compute the baryonic screening masses with nucleon quantum numbers and its negative parity partner in
thermal QCD with $N_f=3$ massless quarks for a wide range of temperatures, from $T \sim 1$ GeV up to
$\sim 160$ GeV. The computation is performed by Monte Carlo simulations of lattice QCD with $O(a)$-improved
Wilson fermions by exploiting a recently proposed strategy to study non-perturbatively QCD at very high
temperature. Very large spatial extensions are considered in order to have negligible finite volume effects.
For each temperature we have simulated $3$ or $4$ values of the lattice spacing, so as to perform the
continuum limit extrapolation with confidence at a few permille accuracy. The degeneracy of the positive
and negative parity-state screening masses, expected from Ward identities associated to non-singlet axial
transformations, provides further evidence for the restoration of chiral symmetry in the high temperature
regime of QCD. In the entire range of temperatures explored, the baryonic masses deviate from the free theory
result, $3 \pi T$, by $4$--$8\%$. The contribution due to the interactions is clearly visible up to the highest
temperature considered, and cannot be explained by the expected leading behavior in the QCD coupling
constant $g$ over the entire range of temperatures explored.
\end{abstract}


\end{frontmatter}



\section{Introduction}
Quantum Chromodynamics (QCD) under extreme conditions is an area of intense research due to
its fundamental r\^ole in many fields of physics, e.g. the cosmological evolution of the early
universe or the interpretation of the results in relativistic heavy ion collision experiments.

It is well known that, even at very high temperatures, the perturbative approach for studying the
QCD dynamics is limited by the so-called infrared problem~\cite{Linde:1980ts}. On the one hand, finding analytic
solutions to overcome this problem is by itself an interesting area of research which is being
actively pursued, see Refs.~\cite{Braaten:1989mz,Blaizot:2003iq,Andersen:2004fp,Mogliacci:2013mca,Laine:2016hma,Arslandok:2023utm}
and references therein. On the other hand, thanks
to the progress achieved in lattice QCD over the last few years, it became possible to study
thermal QCD non-perturbatively from first principles up to very high temperatures~\cite{Giusti:2016iqr,DallaBrida:2021ddx}.

Building on this progress, the calculation of the Equation of State (EoS) in the SU(3) gauge
theory showed beyond any doubt that the contributions which are computable in
perturbation theory are not enough to explain the non-perturbative result
up to temperatures of at least two orders of magnitude above the critical
one~\cite{Borsanyi:2012ve,Giusti:2016iqr}. More recently the computation of the QCD mesonic screening
masses showed that the known perturbative result cannot explain their values up to temperatures of the
order of the electro-weak scale or so~\cite{DallaBrida:2021ddx}. These results point straight to the fact that,
for a reliable determination of the thermal properties of QCD, a fully non-perturbative treatment of the theory
is required up to temperatures of the order of the electro-weak scale or so.

The purpose of this study is to perform the first non-perturbative calculation of the baryonic screening
masses with nucleon quantum numbers over a wide range of temperatures,
from $T\sim 1$ GeV up to $\sim 160$ GeV. This is achieved by extending to the nucleon sector the strategy proposed
in Ref.~\cite{DallaBrida:2021ddx}.  Technically this is feasible because, at asymptotically large temperatures,
baryonic correlators do not suffer from the exponential depletion of the signal-to-noise ratio as they do at
zero temperature~\cite{Parisi:1983ae,Lepage:1989hd}.

Baryonic screening masses are important properties of the quark-gluon plasma. They characterize the
behaviour at large spatial distances of correlation functions of fields carrying baryonic quantum numbers.
Being the inverse of the correlation lengths, they are related to the response of the plasma when a
baryon (nucleon) is injected into the system. Screening masses are also ideal probes to verify the
restoration of chiral symmetry in QCD in the high temperature regime.

While a rich literature is available on the mesonic spectrum, very few studies have been performed on the baryonic one.
In particular, all lattice calculations, both in the quenched approximation~\cite{PhysRevD.36.2828,GOCKSCH1988334}
and in the full theory~\cite{Gottlieb:1987gz}, have been restricted to very low temperatures and no extrapolation
to the continuum limit has ever been performed, see Refs.~\cite{Datta:2012fz,Rohrhofer:2019yko,Aoki:2020noz}
for more recent efforts on the subject.

This letter is organized as follows. In Section \ref{sec:intro} we introduce the definition of nucleon
screening masses, and we discuss how these quantities can probe chiral symmetry restoration. In
Section \ref{sec:corfcn} we briefly review the strategy that we have used to simulate QCD up to the
electro-weak scale, and we give the definition of the baryonic correlation functions and screening
masses on the lattice. In Section \ref{sec:continuum} the values of the screening masses in the continuum
limit at all temperatures considered are given. In Section \ref{sec:res} we discuss our final results,
and present our conclusions. Various technical details are discussed in several appendices.

\section{Definition of baryonic screening masses\label{sec:intro}}
We are interested in the screening masses related to the fermion fields
\be
\label{eq:op}
    N\, = \, \epsilon^{abc} \left( u^{a T} C\gamma_5 d^b\right)d^c\; , \quad 
\overline{N} \, = \overline{d}^e \left( \overline{d}^f C\gamma_5 \overline{u}^{g T} \right) \epsilon^{feg}\, ,
\ee
where the transposition acts on spinor indices, latin letters indicate color indices,
and $C=i\gamma_0\gamma_2$  is the charge-conjugation matrix. The contraction with the totally anti-symmetric
tensor $\epsilon^{abc}$ guarantees that the nucleon field is a color singlet.

The two-point correlation functions we are interested in are
\be
\label{eq:2pt}
C_{N^\pm} (x_3) \, = \, \int dx_0 dx_1 dx_2 e^{-i\frac{x_0}{L_0}\pi} \braket{\Tr \left[ P_{\pm} N(x) \overline{N}(0)\right]}\, ,
\ee
where $P_\pm = (1\pm \gamma_3)/2$ are the projectors on positive ($N^+$) and
negative ($N^-$) $x_3$-parity states respectively, and the trace is over the
free Dirac indices of the fermion fields in Eq.~(\ref{eq:op}). The integral
in Eq.~(\ref{eq:2pt}) selects the component associated to the Matsubara
frequency $\pi/L_0$, which is the lowest one due to the anti-periodic boundary
conditions of fermion fields in the compact direction of length $L_0$.
The screening masses are defined as
\begin{align}
m_{N^{\pm}} \, = \, -\lim_{x_3\rightarrow \infty} \frac{d}{dx_3} \ln \left[ 
C_{N^\pm}(x_3) \right]\, ,
\end{align}
and they characterize the exponential decay of the two-point correlation
functions at large spatial distances.

The comparison of $m_{N^{+}}$ with
$m_{N^{-}}$ provides a quantitative test of the restoration of chiral symmetry
in the high temperature regime. Indeed when chiral symmetry is not spontaneously broken,
the positive and negative parity correlation functions are equal up to a sign in the
chiral limit, see Eq.~(\ref{eq:WI1}) in \ref{app:A}, and therefore
$m_{N^{+}}=m_{N^{-}}$. This is at variance of the zero temperature case, where the screening
masses $m_{N^+}$ and $m_{N^-}$ correspond to the chiral limit values of the nucleon and of
the $N(1535)$ masses. Due to the spontaneous breaking of chiral symmetry, they differ
by several hundreds of MeV~\cite{Workman:2022ynf}.

\subsection{Shifted boundary conditions}
In the rest of this letter, we define the thermal theory in a moving frame by requiring
that the fields satisfy shifted boundary conditions in the compact
direction~\cite{Giusti:2011kt,Giusti:2010bb,Giusti:2012yj}, while we set periodic
boundary conditions in the spatial directions. The former consist in shifting the fields
by the spatial vector $L_0\, \bsxi$ when crossing the boundary of the compact direction,
with the fermions having in addition the usual sign flip. For the gluon and the quark
fields these boundary conditions read
\bea
  A_\mu(x_0+L_0,\bsx) & = & A_\mu(x_0,\bsx - L_0\bsxi)\; ,\nonumber\\
  \psi(x_0+L_0,\bsx) & = & -\, \psi(x_0,\bsx - L_0\bsxi)\; ,\label{eq:psibcs}\\
  \overline{\psi}(x_0+L_0,\bsx) & = & -\,
  \overline{\psi}(x_0,\bsx - L_0\bsxi)\; .\nonumber
\eea
In the presence of shifted boundary conditions the periodic direction is identified
by the vector $L_0(1,\bsxi)$. As a consequence, a relativistic thermal field theory
in the presence of a shift $\bsxi$ is equivalent to the very same theory with usual
periodic (anti-periodic for fermions) boundary conditions but with a longer extension
of the compact direction by a factor $\gamma^{-1}=\sqrt{1+\vec \xi^2}$~\cite{Giusti:2012yj},
i.e. the standard relation between the temperature and the extension in the compact
direction is modified as $T^{-1}= L_0/\gamma = L_0 \sqrt{1+\vec \xi^2}$. The momentum
projection on the lowest Matsubara frequency has also to be modified accordingly. Thanks
to the rotational symmetry of the theory, we can choose one of the axes to be in the
direction of the shift, and restrict our discussion to the case $\bsxi=(\xi,0,0)$.
If we define $(x_0,x_1)$ the first two coordinates of a point in the system with
temporal extent $L_0$ and shifted boundary conditions, and $(x_0',x_1')$
the corresponding ones in the rotated system with temporal extent
$L_0/\gamma$ and periodic boundary conditions, the coordinates are mapped into each other
by a Euclidean Lorentz transformation
\begin{align}
\label{eq:lor_tr}
    \begin{cases}
        x_0' \, = \, \left( x_0 +\xi x_1 \right)\gamma\\
        x_1' \, = \, \left( x_1 -\xi x_0 \right)\gamma \, .
    \end{cases}
\end{align}
The projection of the baryonic correlation function on the first
Matsubara frequency is then achieved by
\be
\label{eq:2pts}
C_{N^\pm} (x_3) \, = \, \int dx_0 dx_1 dx_2 e^{-i\frac{x_0+\xi x_1}{L_0}\gamma^2\pi}
\braket{\Tr \left[ P_{\pm} N(x) \overline{N}(0)\right]}\, ,
\ee
where at variance with Eq.~(\ref{eq:2pt}), the expectation value is
computed in the presence of shifted boundary conditions\footnote{
We use the same notation for correlation functions with or without shifted
boundary conditions since the precise meaning is clear from the context.}.

\section{Lattice strategy, correlation functions and screening masses\label{sec:corfcn}}
We compute the screening masses in QCD with $N_f=3$ massless
quarks\footnote{Technically it is feasible to simulate massless quarks thanks to the large
spectral gap $\pi T$ induced by the temperature in the spectrum of the Dirac operator.}
at the $12$ temperatures $T_0,\dots,T_{11}$ given in Table~\ref{tab:M_CL},
i.e. for $T$ from about $1$ GeV up to approximately $160$ GeV.

We adopt shifted boundary conditions in the compact direction with $\bsxi=(1,0,0)$
and, in order to extrapolate the results to the continuum limit with confidence, several
lattice spacings are simulated at each temperature with the extension of the compact direction
being $L_0/a=4,6,8,10$ while the length of the spatial directions is always $L/a = 288$.
See Appendices A and B of Ref.~\cite{DallaBrida:2021ddx} for the details on the lattice
actions and for the bare parameters of the simulations.

The key idea for reaching very high temperatures on the lattice with a moderate computational
effort is to determine lines of constant physics by fixing the value of a renormalized coupling defined
non-perturbatively in a finite volume~\cite{Giusti:2016iqr,DallaBrida:2021ddx}. The coupling can
be computed precisely on the lattice for values of the renormalization scale $\mu$
which span several orders of magnitude~\cite{Luscher:1993gh,Brida:2016flw,DallaBrida:2018rfy}.
To make a definite choice, we adopt the definition based on the Schr\"odinger functional
(SF)~\cite{Luscher:1993gh} for the temperatures $T_0,\dots,T_8$ and on the gradient flow
(GF)~\cite{Fritzsch:2013hda,Brida:2016flw,DallaBrida:2016kgh} for $T_9, T_{10}$ and $T_{11}$.
\begin{table}[th]
\centering
\begin{tabular}{|c|c|c|c|}
\hline
 & & & \\[-0.25cm]
$T$ & $T({\rm GeV})$  & $\displaystyle \frac{m_{N^+}}{3\pi T} $ &
 $\displaystyle\frac{m_{N^+}-m_{N^-}}{3\pi T}$\\[-0.25cm]
  & & & \\
\hline
$T_0$ &  165(6)   & 1.047(3)  & 0.0006(4) \\
$T_1$ &  82.3(2.8) & 1.0544(19)  & -0.0001(3) \\
$T_2$ &  51.4(1.7) & 1.0569(28)  & 0.0002(3)\\
$T_3$ &  32.8(1.0) & 1.0583(27)  & 0.0003(4)\\
$T_4$ &  20.6(6) & 1.0596(28)  & -0.0011(4)\\
$T_5$ &  12.8(4) & 1.0662(28)  & 0.0001(4)\\
$T_6$ &  8.03(22)  & 1.068(3)  & 0.0001(6)\\
$T_7$ &  4.91(13)  & 1.075(4)  & 0.0004(9)\\
$T_8$ &  3.04(8) & 1.077(4)  & 0.0003(9)\\
$T_9$ &  2.83(7) & 1.076(4)  & 0.0009(12)\\
$T_{10}$& 1.82(4) & 1.089(4)  & 0.0007(20)\\
$T_{11}$& 1.167(23) & 1.078(6)  & 0.0016(15)\\
\hline
\end{tabular}
\caption{Temperatures considered in this letter together with the best results for
  the nucleon mass, $m_{N^+}$, and the mass difference with its parity partner,
  $m_{N^+} - m_{N^-}$, in the continuum limit.\label{tab:M_CL}}
\end{table}

Once the coupling, e.g. $\bar g^2_{\rm SF}(\mu)$, is known in the continuum limit for
$\mu \sim T$ \cite{Brida:2016flw,DallaBrida:2018rfy}, the theory is renormalized by
fixing its value at fixed lattice spacing $a$ to be
\be
\bar g^2_{\rm SF}(g_0^2, a\mu) = \bar g^2_{\rm SF}(\mu)\; ,\quad  a\mu\ll 1\;.
\ee
This is the condition that fixes the so-called lines of constant physics, i.e.
the dependence of the bare coupling constant $g_0^2$ on the lattice spacing, for
values of $a$ at which the scale $\mu$ and therefore the temperature $T$ can be
easily accommodated. In other words, at different temperatures we renormalize
the theory by imposing the value of the renormalized coupling constant at
different scales or equivalently we impose different renormalization conditions
which, however, define the very same renormalized theory at all temperatures. As
a consequence, at each $T$ the theory can be simulated efficiently at various
lattice spacings without suffering from large discretization errors, and the continuum limit
of the observables can be taken with confidence. All technical details on how the
renormalization procedure is implemented in practice are given in Appendices A and B of
Ref. \cite{DallaBrida:2021ddx}.

\begin{figure}[t!]
\vspace{-0.875cm}
  
\includegraphics[width=.5\textwidth]{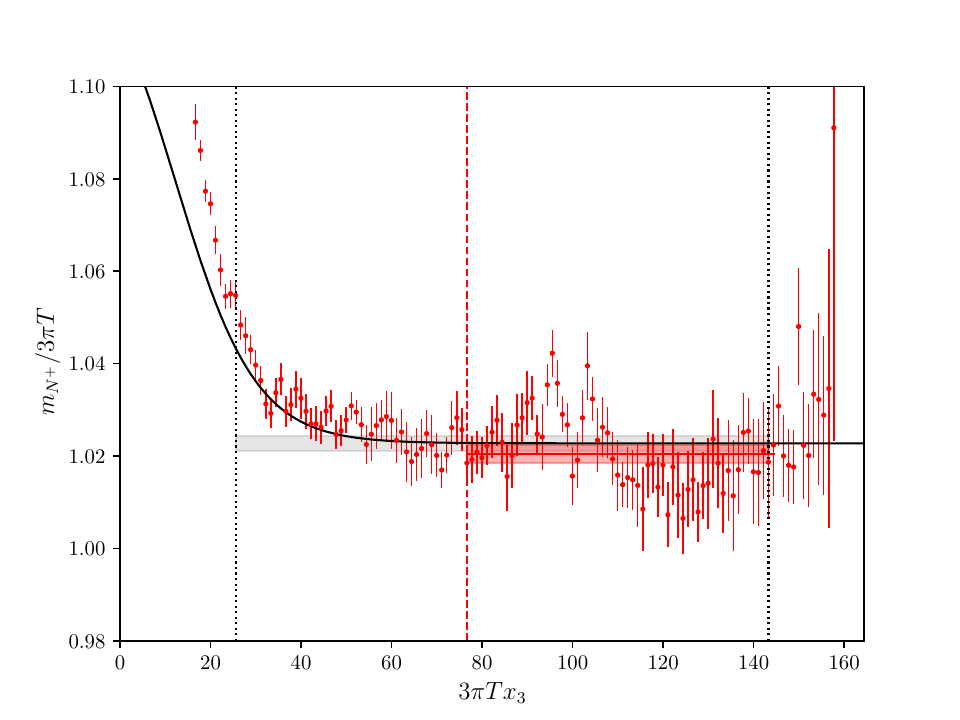}
\caption{Plot of the effective mass $m_{N^{+}}(x_3)$, normalized to $3\pi T$, at
the temperature $T_1$ for $L_0/a = 6$. For the explanation of the various fit
curves and bands see the main text.\label{fig:effmass}}
\end{figure}

The lattice transcription of Eq.~(\ref{eq:2pts}) for $\bsxi=(1,0,0)$, which is the relevant case to this study,
reads\footnote{Even if the use of shifted boundary conditions is not crucial for the calculation of the
screening masses, we have chosen to use them so as to share the cost of generating the gauge configurations
with other projects~\cite{Bresciani:2022lqc,Bresciani:2023zyg}.}
\begin{align}
\label{eq:2ptlat}
    C_{N^\pm} (x_3) \, & = \, a^3 \sum_{x_0,x_1,x_2} e^{-i\frac{x_0+x_1}{2L_0}\pi} \braket{\Tr \left[ P_{\pm} N(x) \overline{N}(0)\right]}\, \nonumber\\
    \, & = \, a^3 \sum_{x_0,x_1,x_2} e^{-i\frac{x_0+x_1}{2L_0}\pi} \braket{\left[ W^1_{\pm}-W^2_{\pm} \right]} \, ,
\end{align}
where the two terms in the second line are the Wick contractions
obtained by integrating over the fermion fields. Their expressions read
\begin{align}
W^1_{N^\pm}\!\! & =\!\! \Tr\left[ S^{ag T}(x,0) C\gamma_5 S^{bf}(x,0)C\gamma_5 \right]\!\!
  \Tr\left[ S^{ce}(x,0) P_{\pm} \right]\epsilon^{abc}\epsilon^{feg}\,,\nonumber\\
W^2_{N^\pm}\!\! & =\!\! \Tr\left[ S^{ag T}(x,0) C\gamma_5 S^{be}(x,0) P_{\pm} S^{cf}(x,0) C\gamma_5 \right] \epsilon^{abc} \epsilon^{feg},
\end{align}
where $S(x,y)$ is the quark propagator of the degenerate quarks.

Once the correlators have been computed, the effective screening masses are defined as
\begin{align}
    m_{N^\pm}(x_3) \, = \, - \frac{1}{a} \ln \left[ \frac{C_{N^\pm} (x_3 + a)}{C_{N^\pm} (x_3)} \right] \, .
\end{align}
As a representative example of the data, the nucleon effective
mass is shown in Fig.~\ref{fig:effmass} for $T_1$ and
$L_0/a =6$. In order to determine the value of the screening mass,
we start by fitting the effective mass to a constant plus a correction
deriving from the contamination of the first excited state (solid black line)
from a minimum value up to the last point where we have a good signal
(black dashed lines). The minimum value is chosen to have a
good quality of the fit and to have, at the same time, a non
vanishing contribution from the first excited state. On one hand,
for the ensembles where the signal is good enough at a large distance,
from this fit we estimate the minimum value $x^{\textrm{min}}_3/a$
(red dashed line) from which the excited state contamination is below
the target statistical precision. The screening mass is then
obtained by averaging the plateau (red band) from $x^{\textrm{min}}_3/a$
up to the last point where we have a good signal. On the other hand, for the lowest
temperatures and for the ensembles corresponding to $L_0/a=10$, where the loss of
signal is more relevant at a large distance, the screening mass is directly estimated
from the results of the effective mass fit (grey band).

Our best estimates of the screening masses are reported in
Tables~\ref{tab:OUTmassesH} and \ref{tab:OUTmassesL} of~\ref{app:B} for all the
lattices simulated. The statistical error varies from a few permille to at most 5 permille
for the smallest temperature. In order to profit from the correlations in our data
for reducing the statistical errors, we also compute $(m_{N^+}-m_{N^-})/(3\pi T)$
and report its values in Tables~\ref{tab:OUTmassesH} and \ref{tab:OUTmassesL} as well.
This combination is particularly interesting because it is a measure of the chiral
symmetry restoration which can be computed very precisely.

We have explicitly checked that finite volume effects are negligible within our statistical
errors: we have generated three more lattices at the highest and at the lowest temperatures
for the smallest spatial volumes corresponding to $L_0/a = 6$, $L_0/a = 10$, and $L_0/a = 8$
for $T_0$, $T_1$, and $T_{11}$ respectively. These lattices have the same dimensions in the
compact and in the $x_3$ directions as those used to extract the results in Tables~\ref{tab:OUTmassesH} and
\ref{tab:OUTmassesL} but smaller extensions in the other two spatial directions. The screening masses
computed on them are in agreement with those calculated on the larger volume, see~\ref{app:B} for details.
Therefore we can safely assume that our results have negligible finite-volume effects within the
statistical precision as expected by the theoretical analysis in Ref.~\cite{DallaBrida:2021ddx}.

\begin{figure}[t!]
\includegraphics[width=0.45\textwidth]{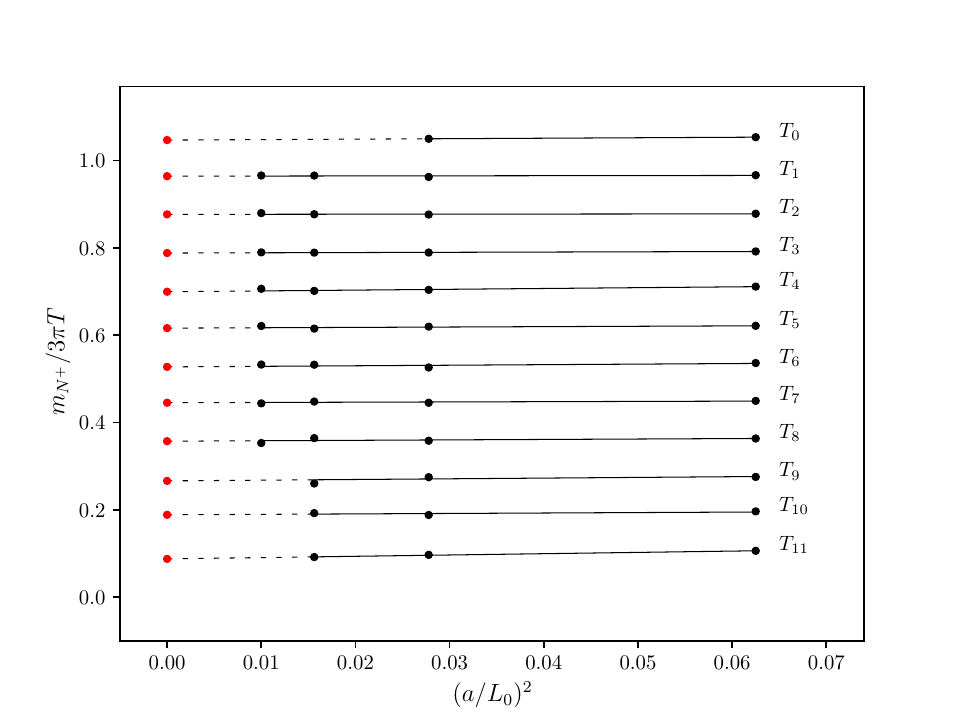}
\caption{Numerical results for the tree-level improved nucleon screening mass at finite
  lattice spacing (black dots, error bars smaller than symbols). The lines in the panel represent
  the linear extrapolations in $(a/L_0)^2$ to the continuum limit. Each temperature is analyzed
  independently from the others. Data corresponding to $T_i$ ($i=0,\dots,11$) are shifted downward
  by $0.09\times i$ for better readability.
\label{fig:MN_extrCL}}
\end{figure}

\section{Continuum limit of baryonic screening masses\label{sec:continuum}}
The results that we have collected at finite lattice spacing have to be extrapolated to the
continuum limit along lines of constant physics. For $O(a)$-improved actions, the Symanzik
effective theory predicts the leading behaviour of the lattice artifacts to be of order $a^2$.
We can accelerate the convergence to the continuum by introducing the tree-level improved
definitions
\be
m_{N^\pm} \longrightarrow m_{N^\pm} -
\Big[m^{\rm free}_{N^\pm} - 3\pi T\Big]\; , 
\ee
where $m^{\rm free}_{N^\pm}$ is the mass in the free lattice theory. As 
shown in the \ref{app:mass_SB}, where the computation is reported, 
the latter is the same for both the $m_{N^+}$ and the $m_{N^-}$  masses.
From now on we will consider always the tree-level improved definition of
the screening masses and indicate them with $m_{N^\pm}$. 

All data for the improved nucleon screening mass are represented in Fig.~\ref{fig:MN_extrCL} where,
in order to improve the readability, data corresponding to $T_i$ ($i=0,\dots,11$) are shifted downward
by $0.09\times i$. At each temperature, lattice artifacts are well described by a single correction
proportional to $(a/L_0)^2$. Indeed by fitting each data set linearly in $(a/L_0)^2$, the values of
$\chi^2/{\rm dof}$ are all around 1 with just a few outliers which, however, are not surprising given
the large amount of data and fits. The results of the fits are shown in Fig.~\ref{fig:MN_extrCL} as
straight lines. For the mass difference $(m_{N^+} - m_{N^-})$, the coefficient of $(a/L_0)^2$ is found to
be compatible with zero at all temperatures. We take the continuum limit values from these fits as our
best results for the nucleon screening mass and for the difference $(m_{N^+} - m_{N^-})$. They are reported
in Table~\ref{tab:M_CL} for all the $12$ temperatures considered. As a further check of the extrapolations,
we have fitted the data by excluding the coarsest lattice spacing, i.e. $L_0/a=4$, for the temperatures
$T_1,\ldots, T_{8}$ for which we have $4$ data points. The intercepts are in excellent agreement with those
of the previous fits, albeit with a slightly larger error. For the same sets of data, we have also attempted
to include in the fit a $(a/L_0)^2\ln(a/L_0)$ or a $(a/L_0)^3$ term. The resulting coefficients are compatible
with zero. Given the high quality of the fits and of the data, it is not necessary to model the temperature
dependence of the discretization effects so as to perform a global fit of the data.
\begin{figure}[t!]
\includegraphics[width=.45\textwidth]{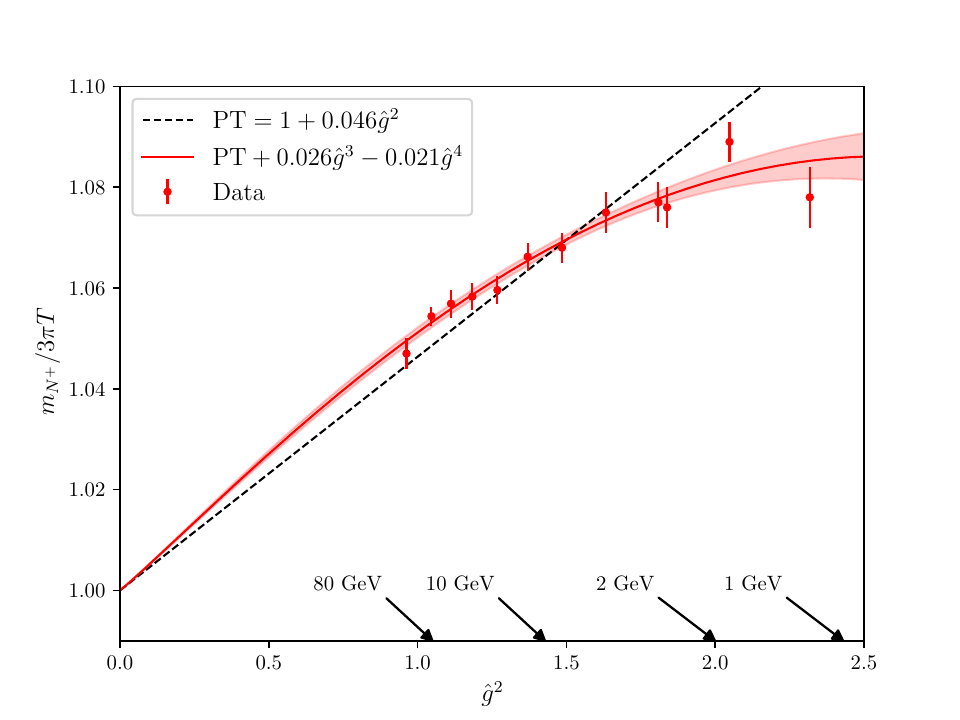}
\caption{Nucleon screening mass versus $\hat g^2$. The band represent the best fit
  to Eq.~(\ref{eq:quartic}), while the dashed line is the analytically known
  contribution in Eq.~\ref{eq:PT}.\label{fig:massCL}}
\end{figure}

\section{Discussion and conclusions \label{sec:res}}
The main results of this paper are the baryonic screening masses reported in
Table~\ref{tab:M_CL}. They have been computed in a wide temperature range starting
from $T\!\sim\! 1$~GeV up to $\sim\! 160$~GeV or so with a precision of a few permille.

The first observation is that, as anticipated in section~\ref{sec:corfcn},
within our rather small statistical errors we find an excellent agreement between
$m_{N^+}$ and $m_{N^-}$. This is a clear manifestation of the restoration of chiral
symmetry occurring at high temperature in line with the analogous results for mesonic
screening masses in Ref.~\cite{DallaBrida:2021ddx}. For this reason in the following
we discuss the nucleon mass $m_{N^+}$ only.

A second observation is that the bulk of the nucleon screening mass is
given by the free-theory value, $3\pi T$, plus $4-8\%$ positive contribution over the
entire range of temperatures explored.

To scrutinize in detail the temperature
dependence induced by the non-trivial dynamics, we introduce the function $\hat g^2 (T)$
defined as
\begin{equation}\label{eq:gmu}
  \frac{1}{\hat g^2(T)} \equiv \frac{9}{8\pi^2} \ln  \frac{2\pi T}{\Lambda_{\MSbar}} 
  + \frac{4}{9 \pi^2} \ln \left( 2 \ln  \frac{2 \pi T}{\Lambda_{\MSbar}}  \right)\; , 
\end{equation}
where $\Lambda_{\MSbar} = 341$~MeV is taken from Ref.~\cite{Bruno:2017gxd}.
It corresponds to the 2-loop definition of the strong coupling constant in
the $\MSbar$ scheme at the renormalization scale $\mu=2\pi T$. For our purposes, however,
this is just a function of the temperature $T$, suggested by the effective theory, that
we use to analyze our results, see Ref.~\cite{DallaBrida:2021ddx} for more details.

The screening masses versus $\hat g^2(T)$ are plotted in Fig.~\ref{fig:massCL}. The dashed line in
this plot is the next-to-leading contribution to the nucleon screening mass which has been computed in
the effective theory only very recently~\cite{Giusti:PT,Hansson:1994nb}.
For three massless quarks, the expression reads
\be
\label{eq:PT}
\frac{m^{\rm nlo}_{N^+}}{3\pi T} = 1 + 0.046 g^2\; ,
\ee
where $g$ is the QCD coupling constant. It is rather clear that from $T=T_0\sim 160$ GeV down to
$T=T_7\sim 5$~GeV the perturbative expression is within half a percent or so with respect to the
non-perturbative data. If a quick convergence of the perturbative series is assumed, this result
would suggest that the bulk of the contribution due to the interactions is given by the $O(g^2)$
term. The full set of data, however, shows a distinct negative curvature which
requires higher orders in $\hat g^2$ to be parameterized. We thus fit the values of $m_{N^+}$
reported in the third column of Table~\ref{tab:M_CL} to a quartic polynomial in
$\hat g$ of the form
\be\label{eq:quartic}
\frac{m_{N^+}}{3\pi T} = b_0 + b_2\, \hat g^2 + b_3\, \hat g^3 + b_4\, \hat g^4\; .  
\ee
The intercept $b_0$ turns out to be compatible with $1$, as predicted by the free theory,
within a large error. We thus enforce it to be the free-theory value, $b_0=1$, and we fit
the data again. The coefficient of the $\hat g^2$ term turns out to be compatible with
the theoretical expectation in Eq.~\ref{eq:PT} within again a large uncertainty. We have
thus fixed also this coefficient to its analytical value, $b_2=0.046$, and we have
performed again the quartic fit of the form in Eq.~\ref{eq:quartic}. As a result, we
obtain $b_3=0.026(4)$, $b_4=-0.021(3)$ and ${\rm cov}(b_3,b_4)/[\sigma(b_3) \sigma(b_4)]=-0.99$
with $\chi^2/{\rm dof}= 0.64$. This is indeed the best parameterization of our results
over the entire range of temperatures explored.

For completeness, we notice that if the cubic coefficient is enforced to vanish,
i.e. $b_3=0$, the fit returns $b_2=0.062(3)$, $b_4=-0.011(2)$ and
${\rm cov}(b_2,b_4)/[\sigma(b_2) \sigma(b_4)]=-0.97$ with again an excellent value
of $\chi^2/{\rm dof}= 0.68$. The coefficient $b_2$ , however, turns out to be in
disagreement with the analytic determination. Data can also be fit to the function
in Eq.~\ref{eq:quartic} with $b_0=1$ and $b_4=0$ but again the coefficient $b_2$ would
be in disagreement with the perturbative result.

Finally we observe that it has been
possible to reach a precision on the nucleon screening mass of a few permille because, at
asymptotically large temperatures, baryonic correlators do not suffer from the exponential
depletion of the signal-to-noise ratio.

\section*{Acknowledgement}
We wish to thank Mikko Laine for several discussions on the topic of this paper. We acknowledge
PRACE for awarding us access to the HPC system MareNostrum4 at the Barcelona Supercomputing
Center (Proposals n. 2018194651 and 2021240051) and EuroHPC for the access to the HPC system
Vega (Proposal n. EHPC-REG-2022R02-233) where most of the numerical results presented in this
paper have been obtained. We also thank CINECA for providing us with computer time on
Marconi (CINECA-INFN, CINECA-Bicocca agree-ments). The R\&D has been carried out on the PC
clusters Wilson and Knuth at Milano-Bicocca. We thank all these institutions for the technical
support. This work is (partially) supported by ICSC – Centro Nazionale di Ricerca in High
Performance Computing, Big Data and Quantum Computing, funded by European Union – NextGenerationEU.

\appendix

\section{Ward Identity in the continuum\label{app:A}}
Under the infinitesimal axial non-singlet transformation
\be
    \label{eq:chiral_sym}
    \delta \psi(x) \, = \, i \epsilon \sigma_3 \gamma_5 \psi(x)\,, \quad 
    \delta \overline{\psi}(x) \, = \,i \epsilon \overline{\psi}(x)\gamma_5  \sigma_3 \, ,
\ee
with $\overline{\psi}=(\bar u,\bar d)$ and analogously for $\psi$
and where $\sigma_3$ is the third Pauli matrix acting on the flavour index,
the nucleon fields in Eq.~(\ref{eq:op})
transform as
\be
\delta N (x) = - i \epsilon \gamma_5 N(x)\;, \quad
\delta \overline{N}(x) = - i \epsilon \overline{N}(x) \gamma_5 \;.
\ee
If we consider the composite field
\be
{\cal O}(x,0) = i \Tr \left[ \gamma_5 P_{\pm} N(x) \overline{N}(0)\right]\, , 
\ee
then 
\be
\delta  {\cal O}(x,0) =
\epsilon \Tr \left[P_{\pm} N(x) \overline{N}(0)\right] +
\epsilon \Tr \left[P_{\mp} N(x) \overline{N}(0)\right]\; . 
\ee
In the chiral limit, and if the symmetry is not spontaneously broken,  
it holds $\langle \delta  {\cal O}(x,0) \rangle = 0$ which in turn
implies, see for instance~\cite{Aarts:2017rrl} for a recent derivation, 
\be\label{eq:WI1}
\braket{\Tr \left[P_{+} N(x) \overline{N}(0) \right]} \, = \, -\braket{\Tr
\left[P_{-} N(x) \overline{N}(0) \right]}\, .
\ee

\section{Simulation details and lattice results\label{app:B}}
\begin{table}[th!]
\centering
\begin{tabular}{|c|c|c|c|c|c|}
\hline
&  &  &  &  &  \\[-0.125cm]
$T$ & $L_0/a$ & $n_{\rm mdu}$ & $n_{\rm nsrc}$   &$\displaystyle\frac{m_{N^+}}{3\pi T}$\!\!\! &
$\!\!\displaystyle\frac{m_{N^+}\! - \!m_{N^-}}{3\pi T}\!\!$\\[-0.125cm]
&  &  &  &  &  \\
\hline
\multirow{2}{*} {$T_0$} 
  &  4 & 300 & 4 & 0.9863(15) & 0.0002(3) \\
  &  6 & 390 & 4 & 1.0178(17) & 0.00041(19) \\
\hline                                        
\multirow{4}{*} {$T_1$} 
  &  4 & 300 & 4 & 0.9892(18) & 0.0001(3) \\
  &  6 & 310 & 4 & 1.0204(20) & 0.0002(4) \\
  &  8 & 500 & 4 & 1.0371(18) & -0.00013(23) \\
  & 10 & 500 & 4 & 1.0438(28) & 0.0003(5) \\
\hline                                        
\multirow{4}{*} {$T_2$} 
  &  4 & 300 & 4 & 0.9909(23) & 0.0001(4) \\
  &  6 & 320 & 4 & 1.0242(24) & -0.00017(28) \\
  &  8 & 490 & 4 & 1.0385(30) & 0.00026(29) \\
  & 10 & 500 & 4 & 1.048(5) & 0.0005(6) \\
\hline                                        
\multirow{4}{*} {$T_3$} 
  &  4 & 300 & 4  & 0.9945(25) & 0.0006(4) \\
  &  6 & 340 & 4 & 1.027(3) & 0.0002(4) \\
  &  8 & 490 & 4 & 1.0406(23) & 0.0005(3) \\
  & 10 & 500 & 4 & 1.048(6) & 0.0003(7) \\
\hline                                        
\multirow{4}{*} {$T_4$} 
  &  4 & 440  & 4 & 1.0040(16) & 0.0007(5) \\
  &  6 & 310 & 4 & 1.0317(26) & -0.0007(4) \\
  &  8 & 490 & 4 & 1.0430(29) & -0.0001(4) \\
  & 10 & 500 & 4 & 1.054(5) & -0.0013(6) \\
\hline                                        
\multirow{4}{*} {$T_5$} 
  &  4 & 310  & 4 & 1.004(3) & -0.0007(6)\\
  &  6 & 310 & 4 & 1.038(3) & 0.0005(8) \\
  &  8 & 500 & 4 & 1.0466(26) & -0.0001(3) \\
  & 10 & 500 & 4 & 1.059(4) & -0.0001(5) \\
\hline                                        
\multirow{4}{*} {$T_6$} 
  &  4 & 300 & 4  & 1.0089(25) & -0.0006(9) \\
  &  6 & 320 & 4 & 1.034(3) & -0.0002(7) \\
  &  8 & 500 & 4 & 1.054(4) & -0.0002(5) \\
  & 10 & 500 & 4 & 1.061(6) & 0.0004(10) \\
\hline                                        
\multirow{4}{*} {$T_7$} 
  &  4 & 320 & 4  & 1.012(4) & 0.0005(12) \\
  &  6 & 310 & 4  & 1.043(4) & 0.0006(7) \\
  &  8 & 500 & 4  & 1.059(3) & -0.0001(8) \\
  & 10 & 500 & 4  & 1.062(6) & 0.0026(17) \\
\hline                                        
\multirow{4}{*} {$T_8$} 
  &  4 & 320 & 8 & 1.016(4) & 0.0023(14) \\
  &  6 & 300 & 8 & 1.046(4) & -0.0001(11) \\
  &  8 & 500 & 4 & 1.066(4) & -0.0007(8) \\
  & 10 & 500 & 5 & 1.061(4) & 0.0013(13)  \\
\hline
\end{tabular}
\hfill
\caption{
Results for the nucleon screening mass, $m_{N^+}$ ,
and the mass difference with its parity partner, $(m_{N^+} - m_{N^-})$,
normalized to $3\pi T$ at finite lattice spacing for the temperatures
$T_0, \ldots, T_8$. The number of molecular dynamic units (MDUs) generated, $n_{\rm mdu}$,
and the number of local sources per configuration on which the two-point correlation functions
have been computed, $n_{\rm nsrc}$, are also reported. The latter are always calculated by
skipping $n_{\rm skip}=10$ MDUs between two consecutive measurements.}\label{tab:OUTmassesH}
\end{table}

\begin{table}[t!]
\centering
\begin{tabular}{|c|c|c|c|l|l|}
  \hline
  &  &  &  &  &  \\[-0.125cm]
  $T$ & $L_0/a$ & $n_{\rm mdu}$ & $n_{\rm nsrc}$   &$\;\;\;\displaystyle\frac{m_{N^+}}{3\pi T}$\!\!\!
  & $\!\!\displaystyle\frac{m_{N^+}\! - \!m_{N^-}}{3\pi T}\!$\\[-0.125cm]
  &  &  &  &  &  \\
\hline
\multirow{3}{*} {$T_9$} 
  &  4 & 400 & 4 & 1.0180(26) & -0.0008(12) \\
  &  6 & 390 & 4 & 1.0526(28) & 0.0005(10) \\
  &  8 & 390 & 4 & 1.052(5) & 0.0002(10) \\
\hline
\multirow{3}{*} {$T_{10}$} 
  &  4 & 410 & 4 & 1.029(4) & -0.0019(21) \\
  &  6 & 400 & 4 & 1.056(3) & -0.0021(14) \\
  &  8 & 390 & 4 & 1.074(3) & 0.0013(17) \\
\hline
\multirow{3}{*} {$T_{11}$} 
  &  4 & 400 & 4 & 1.029(4) & 0.0001(21) \\
  &  6 & 390 & 4 & 1.055(6) & -0.0015(17) \\
  &  8 & 390 & 4 & 1.063(5) & 0.0016(11) \\
\hline 
\end{tabular}
\hfill
\caption{As in Table~\ref{tab:OUTmassesH} but for $T_9$, $T_{10}$ and $T_{11}$.\label{tab:OUTmassesL}}
\end{table}

We have simulated three-flavour QCD as described in Appendix E of Ref.~\cite{DallaBrida:2021ddx}.
We have accumulated a certain number of configurations for the computation of the
EoS~\cite{Bresciani:2023zyg}. Among those, we
have selected some that we have used for the computation of the screening masses.
In particular in Tables ~\ref{tab:OUTmassesH} and \ref{tab:OUTmassesL} we report the number of MDUs
considered and the
number of local sources per configuration on which the two-point correlation functions have been computed.
For each configuration, the best estimates of $C_{N^\pm} (x_3)$ in Eq.~(\ref{eq:2ptlat}) have been obtained
by properly averaging their values from all local sources. The screening masses have then been extracted
as described in Section~\ref{sec:corfcn}. The results are reported in Tables~\ref{tab:OUTmassesH} and
\ref{tab:OUTmassesL} for the 9 highest temperatures $T_0, \ldots$, $T_8$ and for the lowest ones,
$T_9$, $T_{10}$ and $T_{11}$ respectively. 

To explicitly check that finite volume effects are negligible within our statistical errors, we
have generated three more lattices at $T_0$ ($L_0/a=6$), $T_1$ ($L_0/a=10$) and $T_{11}$ ($L_0/a=8$)
at three smaller spatial volumes, namely $6\times 144^2 \times 288$, $10\times 120^2 \times 288$,
and $8\times 144^2 \times 288$ (direction 3 the longest) respectively. On these lattices we have
computed the screening masses. They turn out to be in very good agreement with the analogous ones reported
in Tables~\ref{tab:OUTmassesH} and~\ref{tab:OUTmassesL}, and therefore they confirm the
theoretical expectations that finite volume effects are negligible.

\section{Signal-to-noise ratio for baryon correlation functions at finite temperature \label{app:D}}

\begin{figure}[t!]
\includegraphics[width=.50\textwidth]{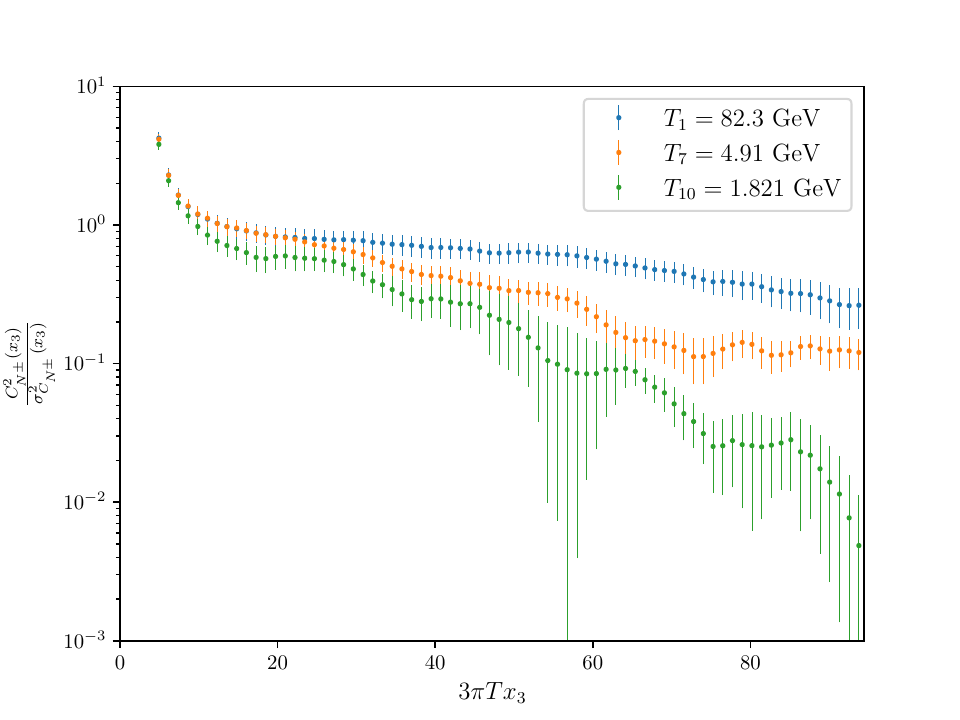}
\caption{Signal-to-noise ratio squared at $T= T_1$, $T_7$ and $T_{10}$
for $L_0/a=6$.\label{fig:NtS}}
\end{figure}
The signal-to-noise ratio squared of baryonic correlation functions goes as 
\be
\frac{C^2_{N^\pm} (x_3)}{\sigma^2_{C_{N^\pm}} (x_3)}\propto \exp\{-(2 m_{N^\pm}-3 m_P) x_3\}+ \dots
\ee
where $\sigma^2_{C_{N^\pm}} (x_3)$ is the variance of $C_{N^\pm}$ and the
$\dots$ stand for sub-leading exponential contributions. At asymptotically
large temperatures, and up to discretization errors, $m_{N^\pm}\rightarrow 3\pi T$ and
$m_P\rightarrow 2\pi T$. As a consequence no exponential depletion of the
signal-to-noise ratio with $x_3$ is expected when $T\rightarrow\infty$. This is at variance of what happens
for $T\rightarrow 0$, where there is a severe
exponential degradation of the signal-to-noise ratio because, for instance,
$(2 m_{N^+}-3 m_P)\rightarrow(2 m_N-3 m_\pi)$ with $m_N$
and $m_\pi$ being the nucleon and the pion masses.

In Fig.~\ref{fig:NtS} it is shown the signal-to-noise ratio squared for
$C_{N^+}$ as a function of $(3\pi T x_3)$ for $3$ temperatures considered in this
letter. As expected the depletion of the signal-to-noise ratio with $x_3$
becomes more severe when the temperature is lowered, and the estimate of the
correlation function becomes noisier. Similar considerations apply to $C_{N^-}$.

\section{Baryonic screening masses in the free lattice theory\label{app:mass_SB}}
In order to accelerate the extrapolation to the continuum limit,
we have computed the baryonic screening masses in the free theory
on the lattice. In the infinite spatial volume limit, the baryonic
correlation functions can be written in the form
\be
C_{N^\pm}(x_3) = \pm \sum_{p_0,q_0}\!\! \int dp_1 dp_2 dq_1 dq_2 \mathcal{M}(p,q)
e^{-2 \,\Omega(p,q,k) x_3}\, ,
\ee
where for shift vectors of the form $\bsxi=(\xi,0,0)$
\begin{equation}\label{eq:MOMferm1}
p_0 = (2 n_0 +1)\frac{\pi}{ L_0}- p_1 \xi\; ,  
\qquad n_0=0,\ldots,L_0/a-1\; , 
\end{equation}
analogously for $q_0$, and the spatial momenta are
$p_k \in [-\pi/a,\pi/a)$. The function $\Omega(p,q,k)$ is given by
\be
\Omega(p,q,k)\,=\,\hat{\omega}(p) + \hat{\omega}(q) + \hat{\omega}(k)\; 
\ee
where $\hat{\omega}$ encodes each quark line contribution to the screening correlator
as defined in Appendix F in Ref. \cite{DallaBrida:2021ddx}, while the matrix
$\mathcal{M}(p,q)$ is a calculable function of the momenta which does
not play a r\^ole in the computation of the baryonic screening mass.
For the lowest Matsubara frequency and for shift vectors of the
form $\bsxi=(\xi,0,0)$, the energy-momentum conservation implies 
\begin{align}
    \begin{cases}
        p_0+q_0+k_0\,=\, \frac{\pi}{L_0} \gamma^2\\
        p_1+q_1+k_1\,=\, \frac{\pi}{L_0}\xi \gamma^2\\
        p_2+q_2+k_2\,=\, 0
    \end{cases}\; .
\end{align}
The screening mass is obtained by minimizing $\Omega(p,q,k)$ with respect
to the momenta. For the shift vector $\bsxi=(1,0,0)$, the minimum is attained
at
\be
\displaystyle p = q = \frac{\pi}{2 L_0} (1,1,0)\; , \qquad k=\frac{\pi}{2 L_0} (-1,-1,0)\; .
\ee
Notice that, since $\hat{\omega}$ is
an even function of the momenta, each quark line gives the same contribution to $\Omega$.
In Table \ref{tab:TL} we list the value of the screening mass
normalized to $3\pi T$ for the temporal extents relevant to this work.
\begin{table}[h]
    \centering
    \begin{tabular}{|c|c|}
        \hline
        $L_0/a$ & $m_{N^\pm}^{\textrm{free}}/3\pi T$ \\
        \hline
        4 & 0.932614077...\\
        6 & 0.967811412...\\
        8 & 0.981401809...\\
        10 & 0.987944825...\\
        \hline
    \end{tabular}
    \caption{Tree-level values of the baryonic screening mass on lattices with temporal extent
      $L_0/a$, infinite spatial volume and shift vector $\bsxi=(1,0,0)$}
    \label{tab:TL}
\end{table}

\bibliographystyle{elsarticle-num} 
\bibliography{bibfile}

\end{document}